\documentclass[dvips]{article}
\usepackage{graphicx}
\usepackage{amssymb}
\usepackage{graphicx}
\usepackage{dcolumn}
\usepackage{amsmath}
\usepackage{lscape}
\usepackage{longtable}

\begin{document}
\textwidth 16cm
\newcommand{\bd}{\begin{document}}
\newcommand{\ed}{\end{document}}
\newcommand{\bc}{\begin{center}}
\newcommand{\ec}{\end{center}}
\newcommand{\bfr}{\begin{flushright}}
\newcommand{\efr}{\end{flushright}}
\newcommand{\lt}{\left}
\newcommand{\rt}{\right}
\newcommand{\vs}{\vspace}
\newcommand{\hs}{\hspace}
\newcommand{\beq}{\begin{equation}}
\newcommand{\eeq}{\end{equation}}
\newcommand{\lb}{\linebreak}
\newcommand{\pb}{\pagebreak}
\newcommand{\mb}{\makebox}
\newcommand{\fb}{\framebox}
\newcommand{\mc}{\multicolumn}
\newcommand{\ben}{\begin{enumerate}}
\newcommand{\een}{\end{enumerate}}
\newcommand{\bit}{\begin{itemize}}
\newcommand{\eit}{\end{itemize}}
\newcommand{\ol}{\overline}
\newcommand{\un}{\underline}
\newcommand{\lefq}{\lefteqn}
\newcommand{\ba}{\begin{array}}
\newcommand{\ea}{\end{array}}
\newcommand{\beqa}{\begin{eqnarray}}
\newcommand{\eeqa}{\end{eqnarray}}
\newcommand{\beqas}{\begin{eqnarray*}}
\newcommand{\eeqas}{\end{eqnarray*}}
\newcommand{\bfg}{\begin{figure}}
\newcommand{\efg}{\end{figure}}
\newcommand{\bds}{\begin{displaymath}}
\newcommand{\eds}{\end{displaymath}}
\newcommand{\btb}{\begin{tabbing}}
\newcommand{\etb}{\end{tabbing}}
\newcommand{\para}{\parallel}
\newcommand{\pad}{\partial}
\newcommand{\nn}{\nonumber}
\newcommand{\la}{\leftarrow}
\newcommand{\ra}{\rightarrow}
\newcommand{\lgla}{\longleftarrow}
\newcommand{\lgra}{\longrightarrow}
\newcommand{\La}{\Leftarrow}\newcommand{\Ra}{\Rightarrow}
\newcommand{\Lra}{\Leftrightarrow}
\newcommand{\Lgla}{\Longleftarrow}
\newcommand{\Lgra}{\Longrightarrow}
\newcommand{\bm}{\boldmath}
\newcommand{\lan}{\langle}
\newcommand{\ran}{\rangle}
\renewcommand{\a}{\alpha}
\renewcommand{\b}{\beta}
\newcommand{\g}{\gamma}
\newcommand{\G}{\Gamma}
\renewcommand{\d}{\delta}
\newcommand{\eps}{\epsilon}
\newcommand{\Th}{\Theta}
\newcommand{\s}{\sigma}
\newcommand{\lam}{\lambda}
\newcommand{\D}{\Delta}
\newcommand{\vare}{\varepsilon}
\newcommand{\pr}{\prime}
\newcommand{\ro}{\rho}
\newcommand{\nab}{\nabla}
\newcommand{\m}{\mu}
\newcommand{\n}{\nu}
\newcommand{\Sg}{\Sigma}
\newcommand{\p}{\pi}
\newcommand{\R}{I\!\!R}
\newcommand{\om}{\omega}
\newcommand{\Om}{\Omega}
\newcommand{\ze}{\zeta}
\newcommand{\vart}{\vartheta}
\newcommand{\tri}{\triangle}
\newcommand{\f}{\frac}
\newcommand{\iny}{\infty}
\newcommand{\pro}{\propto}
\bc {\huge Factorization and Lie Point Symmetries of General Lienard-type Equation in the Complex Plane} \ec

\vs{1cm}

\bc
{\it \"Ozlem Ye\c{s}ilta\c{s}\dag \ddag{\footnote {e-mail : yesiltas@gazi.edu.tr}\\
\dag Department of Mathematics and Statistics, Concordia University
1455 de Maisonneuve Boulevard West, Montr´eal
Qu´ebec, Canada H3G 1M8
\\
\ddag Department of Physics, Faculty of Arts and Sciences,
Gazi University,
06500 Ankara, Turkey\\
\vspace{.15cm} }} \ec
\vs{4cm}

\begin{abstract}
We present a variational approach to general Lienard type equation in order to linearize it and  as an example  Van der Pol oscillator is discussed. The new equation which is almost linear is factorized. The point symmetries of the deformed equation is also discussed and the two dimensional Lie algebraic generators are obtained.\\
{\small \sl \noindent Key words: Lienard, symmetry  \\[0.2cm]
}
\end{abstract}
\baselineskip 0.9cm
\newpage
\section{Introduction}
Investigation of partial and ordinary nonlinear  differential equations is one of the popular recent problems in mathematical physics.  A typical example of a relaxation oscillator is known as Van der Pol limit cycle \cite{1}. Van der Pol equation can be adapted to study the Lienard equation. This paper mainly devoted to study Lienard type equations whose dynamic properties have been widely studied due to the applications in many fields such as physics, mechanics and engineering \cite{1,2,3,4,5,6,7,13}. This kind of general equations and their properties are studied within many methods. Factorization method \cite{infeld} is one of these topics and it can be applied to a class of nonlinear differential equations including  commonly Lienard type equations \cite{8,9,10}. Beyond the factorization method, Lie group approach which is an integrability method is a powerful tool for obtaining solutions and investigating the symmetry properties of the differential equations  \cite{ibra, gaz, ay}. In this paper, variational approach and factorization of the system are discussed in section 2. In addition, section 3 involves a short discussion of the Lie symmetries of the system, the results are concluded in section 4.
\section{Linearizing and factorization}
During the process of deriving Euler's equation one writes
\begin{equation}
    y(x,\alpha)=y(x,0)+\alpha y_{1}(x).
\end{equation}
If we write a difference equation of the form
\begin{eqnarray}
% \nonumber to remove numbering (before each equation)
  \Delta f &=& f[x, y(x,\alpha), y^{'}(x,\alpha)]-f[x, y(x,0), y^{'}(x,0)] \\
   &=& \left( \frac{\partial f}{\partial y} y_{1}+\frac{\partial f}{\partial y^{'} }y^{'}_{1} \right) \alpha+....
\end{eqnarray}
and then continue with introducing the variation of $f$ as
\begin{equation}
    \delta f=\frac{\partial f}{\partial y} \delta y + \frac{\partial f}{\partial y^{'}} \delta y^{'}
\end{equation}
where $\delta y=\alpha y_{1}$ and $\delta y^{'}=\alpha y^{'}_{1}$. Thus the variation of $f$ is expressed in terms of constant $\alpha$. Thus, this operation can be applied to a differential equation called as general Lienard equation:
\begin{equation}\label{nl}
    \ddot{x}+F(x(t)) \dot{x}+x=0
\end{equation}
where the dots correspond derivatives of $x(t)$ with respect to $t$. Now apply the variational operator to (\ref{nl}) that is defined by $\delta x^{n}= n x^{n-1} \delta x= y^{n}$. Here the differential  and variational operators are commuting such that $\delta \frac{dx}{dt} =\frac{d}{dt} \delta x$. In this condition (\ref{nl}) becomes
\begin{equation}
    \ddot{y}+ F(x(t))\dot{y}+\delta(F(x(t))) \dot{x}+G(t)y=0.
\end{equation}
As an example, we can discuss Van der Pol type equation where $F(x(t))=\varepsilon (1-x(t)^{2})$. In this case (\ref{nl}) will be introduced as a complex differential equation \cite{lopez}
\begin{equation}\label{complex}
    \ddot{z}-\varepsilon(1-|z|^{2})\dot{z}+z=0
\end{equation}
where the complex variable $z=z_{1}+i z_{2}$ and the equation admits the translation symmetry $z(t)\rightarrow z(t+\delta)$, phase symmetry $z(t)\rightarrow e
^{i \theta } z(t)$, spatial symmetry $z(t)\rightarrow -z(t)$, conjugation symmetry $z(t)\rightarrow z^{*}(t)$. (\ref{complex})  can be written in a general form:
\begin{equation}\label{cgen}
    \ddot{z}-\varepsilon(1-|z|^{2})\dot{z}+G(t)z=0.
\end{equation}
where $G(t)$ is an unknown function. If the variational operation is applied to (\ref{cgen})
\begin{equation}\label{varcomp}
    \delta\ddot{z}-\varepsilon(1-|z|^{2})\delta\dot{z}+\varepsilon \dot{z}~ \delta|z|^{2}~ G(t)~ \delta z=0
\end{equation}
is obtained and
(\ref{cgen}) turns into the form
\begin{equation}\label{solu}
    \ddot{Y}+\varepsilon (|z(t)|^{2}-1)\dot{Y}+\varepsilon \dot{z}(t) Y^{2}+G(t) Y=0
\end{equation}
here $\delta z_{1}=Y(t)$ is used, we assume that $\delta z_{2}=0$ in the equation given above, according to non-analytical $zz^{*}$. Henceforth, we can discuss the factorization of (\ref{solu}). We use a homographic transformation $D(\alpha, \beta, \theta)$ that is generally given by
\begin{equation}
    (y,t)\rightarrow (Y,T): y=\frac{\alpha(t) Y+\tilde{\beta}(t)}{\gamma(t)Y+\delta(t)},\,\ \alpha \delta-\tilde{\beta} \gamma \neq 0,\,\,\ T=\theta(t),\,\ \tilde{\beta}(t)=\beta_{0} \beta(t)
\end{equation}
and makes the ordinary differential equation (\ref{solu}) is form invariant. Under this transformation $D(\alpha, \beta, \theta)$, (\ref{solu}) turns into
\begin{equation}\label{solu1}
\begin{split}
    \frac{d^{2}Y}{dT^{2}}&+\frac{1}{\theta^{'}} \left(2\frac{\alpha^{'}}{\alpha}+\frac{\theta^{''}}{\theta}-\varepsilon (x^{2}-1)\right)\frac{dY}{dT}
  \\  -& \varepsilon \dot{x} \frac{\alpha}{\theta^{'2}} Y^{2}+ \left(\frac{\alpha^{''}}{\alpha}+2\varepsilon \dot{x} \tilde{\beta}-\varepsilon (1-x^{2})\frac{\alpha^{'}}{\alpha}\right)\frac{Y}{\theta^{'2}}\\+& \frac{1}{\alpha^{-1}\theta^{'2}}(\tilde{\beta}^{''}+\varepsilon\dot{x}
     \tilde{\beta}^{2}-\varepsilon(1-x^{2}) \tilde{\beta}^{'} -G(t))=0,\,\ \alpha \theta^{'}\neq 0
    \end{split}
\end{equation}
In order to transform (\ref{solu1}) into an equation below that is addressed as canonic form of type I in \cite{14}:
\begin{equation}\label{pain}
    \frac{d^{2}Y}{dT^{2}}-6Y^{2}+6 \beta_{0}=0
\end{equation}
one must write
\begin{eqnarray}\label{must}
% \nonumber to remove numbering (before each equation)
  \theta^{'2} &=&- \frac{\varepsilon \dot{x} \alpha}{6} \\ \nonumber
  (Ln~~\alpha)^{'} &=& - \frac{2}{5} (\varepsilon(1-x^{2})+\frac{\ddot{x}}{2\dot{x}}) \\ \nonumber
  \tilde{\beta} &=& -\frac{1}{2\varepsilon \dot{x}}(-G(t)+\varepsilon (1-x^{2})(Ln~~ \alpha)^{'}+(Ln~~ \alpha)^{''}+(Ln~~ \alpha)^{'2}).
\end{eqnarray}
In \cite{sengul} it is pointed out that (\ref{must}) is essential to obtain (\ref{pain}) and equations in (\ref{must}) must be satisfied. In this study we will follow a more general way, there won't be such a restriction given in (\ref{must}). We will use $F_{1}(t)=\varepsilon (|z(t)|^{2}-1)$, $F_{2}(t)=\varepsilon \dot{z}(t)$ in (\ref{solu}) that can be factorized as
\begin{equation}\label{fac}
    \left(\frac{d}{dt}-f_{2}(Y)\right) \left(\frac{d}{dt}-f_{1}(Y)\right)Y=0
\end{equation}
where $f_{1}$ and $f_{2}$ satisfy the relations:
\begin{eqnarray}\label{must2}
% \nonumber to remove numbering (before each equation)
  F_{1} &=& -\left(f_{1}+f_{2}+Y\frac{df_{1}}{dY}\right) \\ \nonumber
  F_{2}Y+G &=& f_{1}f_{2}.
\end{eqnarray}
We want to show that if (\ref{solu}) admits the factorization for $f_{1}$ and $f_{2}$ have the form
\begin{eqnarray}\label{must1}
% \nonumber to remove numbering (before each equation)
  f_{1} &=& \alpha (\sqrt{F_{2}Y}+i\sqrt{G}) \\ \nonumber
   f_{2} &=& \alpha^{-1} (\sqrt{F_{2}Y}-i\sqrt{G})
\end{eqnarray}
where $\alpha$ is a constant and can be found as $\alpha=\pm i \sqrt{\frac{2}{3}}$ by comparing (\ref{must1}) and (\ref{must2}). Then one can find a relation between $F_{1}$ and $G$ using the first equation in (\ref{must2})
\begin{equation}\label{F1-G}
    F_{1}=\pm \frac{5}{\sqrt{6}}\sqrt{G}.
\end{equation}
The second bracket in (\ref{fac}) allows us to write
\begin{equation}
    \dot{Y}\mp i \sqrt{\frac{2F_{2}}{3}}Y^{\frac{3}{2}}\pm \sqrt{\frac{2G}{3}}Y=0
\end{equation}
which is a Bernoulli differential equation and we obtain the solutions of the form \cite{krey}
\begin{equation}
    Y=-\mu(t)^{2}\left(\int^{t}\sqrt{\frac{F_{2}(t^{'})}{6}}\mu(t^{'})dt^{'}\right)^{-2}, ~~~~\mu~(t)=e^{\mp\int^{t}\sqrt{\frac{G(t^{'})}{6}}dt^{'}}
\end{equation}
In order to derive special exact solutions, assume that $(\partial_{t}-f_{1})Y=\psi(t)$, and it yields
\begin{eqnarray}
% \nonumber to remove numbering (before each equation)
  \dot{Y}-f_{1}(Y)Y &=& \psi(t) \\
  \dot{\psi}-f_{2}(Y) \psi&=& 0
\end{eqnarray}
and this choice will lead to have more general solutions. In \cite{8} the author selected a constant function $f_{2}(Y)$ but we want to use a non-constant $f_{2}$ which is given by the second formula in (\ref{must1}). Then, $(21)$ can be given by
\begin{equation}\label{sl}
    \dot{Y}=f_{1}(Y)Y+e^{\int^{t}f_{2}(Y)dt^{'}}.
\end{equation}
Thus, $Y$ can be derived easily as solving this equation
\begin{equation}\label{solll}
\dot{\omega}\mp \sqrt{\frac{G(t)}{6}} \omega=\mp i \sqrt{\frac{F_{2}}{6}}-\frac{1}{2}\omega^{3}e^{\mp i \sqrt{\frac{3}{2}} \int^{t}\left(\sqrt{\frac{F_{2}}{\omega}}-i\sqrt{G}\right)dt^{'}}
, ~~~~\omega~=Y^{-\frac{1}{2}}.
\end{equation}
The solution is introduced as \cite{krey}
\begin{equation}\label{soln}
    \omega(t)=\pm \frac{i}{\mu(t)} \int^{t} \sqrt{\frac{F_{2}(t^{'})}{6}}\mu(t^{'})dt^{'}-\frac{1}{2\mu(t)} \int^{t} \mu(t^{'})\varphi (t^{'})
    e^{\mp \int^{t} \sqrt{\frac{3}{2}G(t^{'})}dt^{'}}
\end{equation}
where  $t$ dependent function $\varphi$ is defined by
\begin{equation}\label{phii}
    \frac{\dot{\varphi}}{\varphi}=\mp i \sqrt{\frac{3F_{2}(t)}{2\omega(t)}}+\frac{3\dot{\omega}}{\omega}.
\end{equation}
Assume that $\varphi$ is a constant. Thus, a special solution for $\omega$ is given by
\begin{equation}
    \omega(t)=\frac{1}{4}\left(C_{1}\pm i\int^{t} \sqrt{\frac{F_{2}(t^{'})}{6}} dt^{'}\right)^{2}
\end{equation}
where $C_{1}$ is an integration constant. Now, let us choose $F_{2}$ as
\begin{equation}\label{fe2}
    F_{2}(t) = -6 e^{\mp \sqrt{6}\int^{t}\sqrt{G}dt^{'}}
\end{equation}
then, using (\ref{fe2}) and  (\ref{F1-G}), the following expression can be obtained
\begin{equation}\label{xxx}
    \dot{z}=-\frac{6}{\varepsilon} e^{\mp\frac{6\varepsilon}{5}\int^{t}(|z|^{2}-1)dt^{'}}
\end{equation}
and the solution of $z$ is introduced by
\begin{equation}\label{zsol}
    z=c e^{i \theta t}
\end{equation}
where $c$ and $\theta$ are some constants. In this case, (\ref{xxx}) is satisfied when
\begin{eqnarray}\label{tik}
% \nonumber to remove numbering (before each equation)
  \varepsilon &=& i \\
  c(c^{2}-1) &=& \mp 5 \\
   \theta &=& \pm \frac{1-c^{2}}{5}.
\end{eqnarray}
 Thus, $F_{1}$, $F_{2}$ and $G$ are expressed by
 \begin{eqnarray}\label{F12G}
 % \nonumber to remove numbering (before each equation)
   F_{1} &=& i(c^{2}-1) \\ \nonumber
   F_{2} &=& -c \theta e^{i \theta t} \\ \nonumber
   G &=& \frac{6}{25}(1-c^{2}).
 \end{eqnarray}
\section{Symmetries of the deformed equation}
The point symmetry of (\ref{solu}) will be discussed here. As it is stated in $[11]$, the differential equations that forms a system
\begin{equation}
    F_{i}(x,u,u_{(1)},...,u_{(k)})=0, \,\,\ i=1,...,s
\end{equation}
is invariant under the $X$ which is an infinitesimal generator such that
\begin{equation}
    XF_{i}|_{F_{i}=0}=0.
\end{equation}
So, the infinitesimal generator for (\ref{solu}) will have a form
\begin{equation}
    X=\xi(x,y)\frac{\partial}{\partial x}+\eta(x,y) \frac{\partial}{\partial y}
\end{equation}
for a second order equation
\begin{equation}
    y^{''}=f(x,y,y^{'})
\end{equation}
where the point transformations in the plane are given by
\begin{eqnarray}
% \nonumber to remove numbering (before each equation)
  \bar{x} &\approx &  x+a \xi(x,y)\\
  \bar{y} &\approx &  y+a \eta(x,y)
\end{eqnarray}
Thus we have
\begin{equation}
    X(y^{''}-f(x,y,y^{'}))|_{y^{''}=f}=0.
\end{equation}
The equation above can also be written $[13]$
\begin{equation}\label{ib}
\begin{split}
    \eta_{xx}&+(2\eta_{xy}-\xi_{xx})y^{'}+(\eta_{yy}-2\xi_{xy})y^{'2}-y^{'3}\xi_{yy}-\xi f_{x}-\eta f_{y}\\+&
    (\eta_{y}-2\xi_{x}-3y^{'}\xi_{y})f-(\eta_{x}+(\eta_{y}-\xi_{x})y^{'}-y^{'2}\xi_{y})f_{y^{'}}=0
    \end{split}
\end{equation}
In this case, $f=-F_{1}y^{'}-F_{2}y^{2}-Gy$ and here the derivatives are taken with respect to $x$ and $y$ is used instead of $Y$ here. We can group the equations as
\begin{equation}\label{1}
    (y^{'})^{3}:\,\,\ \xi_{yy}=0
\end{equation}
\begin{equation}\label{2}
    (y^{'})^{2}:\,\,\ \eta_{yy}-2\xi_{xy}+2F_{1}\xi_{y}=0
\end{equation}
\begin{equation}\label{3}
    y^{'}:\,\,\ 2\eta_{xy}-\xi_{xx}+\xi F_{1x}+F_{1} \xi_{x}=0
\end{equation}
\begin{equation}\label{4}
    y^{2}:\,\,\ \xi F_{2x}-F_{2}(\eta_{y}-2\xi_{x})=0
\end{equation}
\begin{equation}\label{5}
    y : \xi G_{x}+2F_{2}\eta-G(\eta_{y}-2\xi_{x})=0
\end{equation}
\begin{equation}\label{6}
    y^{0} : \eta_{xx}+\eta_{x}F_{1}-\eta G=0.
\end{equation}
\begin{equation}\label{7}
    y y^{'} : 3G\xi_{y}=0
\end{equation}
Thus we can obtain $\xi$ from (\ref{1})
\begin{equation}\label{chi}
    \xi=\alpha(x)y+\beta(x).
\end{equation}
On the other hand, from (\ref{2}), $\eta$ takes the form
\begin{equation}\label{eta}
    \eta=W(x)y+S(x).
\end{equation}
$W(x)=0$ and $\alpha(x)=0$ can be obtained by using (\ref{5}) and (\ref{7}). If $\eta_{y}-2\xi_{x}$ in (\ref{4}) is used in (\ref{5}) we get
 \begin{equation}
 2F_{2}\eta+\beta(x) \frac{F_{2x}}{F_{2}} G=0
\end{equation}
and using (\ref{3})
\begin{equation}\label{}
    \beta(x)=A+B e^{F_{1}x}
\end{equation}
where $A, B$ are constants. We can obtain $S(x)$ using (\ref{6}) as
\begin{equation}\label{S}
    S(x)=C_{1} e^{-\alpha_{+}x}+C_{2} e^{-\alpha_{-}x}
\end{equation}
where $C_{1}, C_{2}$ are constants and $\alpha_{\pm}$ can be given in terms of the parameters of (\ref{solu}):
\begin{equation}\label{al}
    \alpha_{\pm}=\frac{1}{2} \left(F_{1}\pm \frac{\sqrt{F^{2}_{1}+4G}}{2}\right).
\end{equation}
If we use the values of $F_{1}$, $F_{2}$ and $G$ and (\ref{5}) we have
\begin{equation}
    2F_{2}\eta-G(\eta_{y}-2\xi_{x})=0
\end{equation}
where the derivative of $G$ is zero as it is seen from (\ref{F12G}). Then, we can obtain
\begin{equation}
    \alpha_{+}=-i\theta+F_{1},~~~~\alpha_{-}=i\theta
\end{equation}
Finally we get the value of $\theta$ which is same with the result obtained in the previous section
\begin{equation}\label{tita}
    \theta=\frac{1-c^{2}}{5}
\end{equation}
which means that the positive sign should be used for $\theta$ in $(33)$. Thus, two dimensional generators are obtained as
\begin{eqnarray}
% \nonumber to remove numbering (before each equation)
  X_{1} &=& \frac{\partial}{\partial x}+e^{-\alpha_{+}x} \frac{\partial}{\partial y} \\
  X_{2} &=& e^{F_{1}x}\frac{\partial}{\partial x}+e^{-\alpha_{-}x}\frac{\partial}{\partial y}
\end{eqnarray}
the commutation between these operators is $[X_{1},X_{2}]\neq 0$ and the group is not closed.
\begin{figure}
\begin{center}
\includegraphics{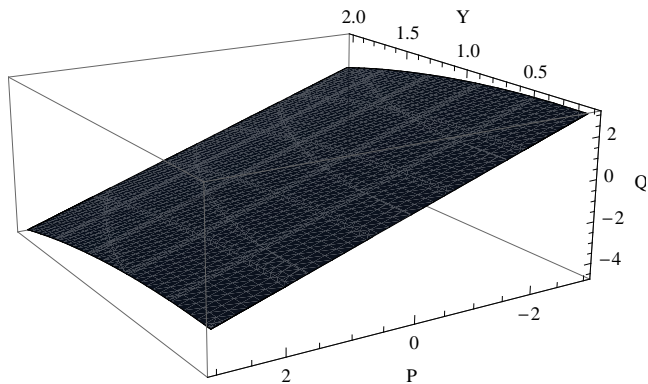}
\caption{The skeleton of (\ref{solu}) for some specific parameters. $\dot{Y}=P,~~ \ddot{Y}=Q$. $\mathcal{M}=\{Y, \dot{Y}, \ddot{Y}\} $ reduced manifold.}
\label{label}
\end{center}
\end{figure}

\section{Conclusion}

The flexibility of the variational approach is shown by applying the variational operator to the Lienard type equation and linearized Van der Pol differential equation is examined in the complex plane. The deformed differential equation is factorized and solutions are obtained. Moreover, it is shown that the time dependent coefficients of the linearized equation that are, $F_{1}$ and $F_{2}$, found as pure imaginary and complex for the real values of $c$. Thus, interesting result is the parameter $\varepsilon$ is obtained as $i$. It means that the symmetry properties of the equation can be changed under the time reversal symmetry within the contributions of  $F_{1}$, $F_{2}$ and $G$ such that the original equation (\ref{complex}) is invariant under spatial reversal but (\ref{solu}) is not invariant under $Y\rightarrow -Y$ and it is not invariant under both spatial and $\varepsilon=i$, $i\rightarrow -i$ parameter and $t\rightarrow -t$ reversal. Besides factorization, the symmetry of equation (\ref{solu}) is examined within point symmetries and two dimensional generators are obtained. The group is not closed, however we have concluded the parameter $\theta$ of $z(t)$ is the same as that obtained with both the symmetry
equations and factorization results.

The skeleton for $(10)$ is a flat surface such that this
equation has two-dimensional generators, as is shown in
figure 1.

\end{document}